\documentclass[useAMS,usenatbib]{mn2e}

\usepackage{amssymb}
\usepackage{amsfonts}
\usepackage{mncite}
\usepackage{epsfig}
\usepackage{psfig}

\newcommand{\msunyr}{M_{\sun}/\mbox{yr}}

\defcitealias{kenyon91}{KH91}

\begin{document}

\title[Photometric evolution of FU Orionis]{The photometric evolution
of FU Orionis objects: disc instability and wind-envelope interaction}

\author[C. Clarke, G. Lodato, S. Y. Melnikov and
M. A. Ibrahimov]{C. Clarke$^1$, G. Lodato$^1$, S. Y. Melnikov$^2$ and
M. A. Ibrahimov$^2$\\ $^1$ Institute of Astronomy, Madingley Road,
Cambridge, CB3 0HA\\ $^2$ Ulugh Beg Astronomical Institute, Academy of
Sciences of Uzbekistan, Astronomical str. 33, Tashkent 700051,
Uzbekistan}

\maketitle

\begin{abstract}

We present the results of a photometric monitoring campaign of three
well studied FU Orionis systems (FU Orionis, V1057 Cygni and V1515
Cygni) undertaken at Maidanak Observatory between 1981 and 2003. When
combined with photometric data in the literature, this database
provides a valuable resource for searching for short timescale
variability - both periodic and aperiodic - as well as for studying the
secular evolution of these systems. In the case of V1057 Cyg (which is
the system exhibiting the largest changes in brightness since it went
into outburst) we compare the photometric data with time dependent
models. We show that prior to the end of the `plateau' stage in 1996,
the evolution of V1057 Cyg in the $V-(B-V)$ colour-magnitude diagram
is well represented by disc instability models in which the outburst is
triggered by some agent - such as an orbiting planet - in the inner
disc.  Following the end of the plateau phase in 1996, the dimming and
irregular variations are consistent with occultation of the source by a
variable dust screen, which has previously been interpreted in terms of
dust condensation events in the observed disc wind. Here we instead
suggest that this effect results from the interaction between the wind
and an infalling dusty envelope, the existence of this envelope having
been previously invoked in order to explain the mid infrared emission
of FU Orionis systems. We discuss how this model may explain some of
the photometric and spectroscopic characteristics of FU Orionis systems
in general.

\end{abstract}


\section{Introduction}

Despite being a rather small class of young stellar objects, the
outbursting FU Orionis systems have been studied extensively over the
last 20 years, since they represent an interesting ``laboratory'' to
study the accretion process on newly born stars and the interaction
between the protostar and its environment. In fact, according to the
most popular interpretation \citep{hartmann96}, their outbursts are
associated with episodes of enhanced accretion through an otherwise
``normal'', T Tauri like, protostellar disc (see, however,
\citealt{herbig03} for a dissenting view). Support to this
interpretation comes mainly from the modeling of the broad band
Spectral Energy Distribution (SED) in terms of accretion disc SEDs,
from their apparent spectral type, which is characterized by high
effective temperatures at smaller wavelength and by lower effective
temperatures at larger wavelengths, and from the shape of some optical
and near-infrared absorption lines, which show the typical double peak
expected from a rotating disc \citep{hartmann85,hartmann87,kenyon88}.

The time evolution of the outbursts may vary considerably from object
to object. If we refer to the three best studied objects (FU Ori
itself, V1057 Cyg and V1515 Cyg, the only ones for which a detailed
light curve has been recorded since the beginning of the outbursts),
it can be seen that while for some of them (FU Ori and V1057 Cyg) the
transition to the outburst phase is rather abrupt (with rise timescale
of the order of 1 yr), for V1515 Cyg it is considerably more gentle
(rise timescale $\approx 20$ yrs). The post-outburst light curve is
also quite different for different objects: while FU Ori has remained
almost steadily in the high luminosity state for at least 70 yrs,
V1057 Cyg has faded rapidly (with a timescale of roughly 10 yrs) and
then showed a ``plateau'' in the light curve, until eventually
undergoing an abrupt luminosity drop in the mid 90s, followed by a
rather erratic photometric variability. V1515 Cyg, on the other hand
(which is characterized by a smaller peak luminosity with respect to
the other two), while remaining overall in the high luminosity state,
has often shown strong photometric variability. It would then appear
to be difficult to reconcile this wide variety of behaviours within a
single model. 

However, the differences in rise timescales can be understood when one
considers the detailed outburst mechanism. This is generally considered
to be due to the onset of a thermal instability in a protostellar disc
fed at a high enough rate \citep{bellin94}, a mechanism analogous to
the one considered in dwarf novae outbursts. In this picture, partial
ionization of hydrogen in the inner disc gives rise to a thermal
instability and to a limit cycle behaviour, that can be able to
reproduce repetitive outbursts. It can be shown \citep{LPF85} that
outbursts initiated at the inner edge of the disc propagate inside-out,
resulting in a long rise timescale, while outbursts triggered somehow
far from the inner edge are characterized by short rise timescales. It
is then possible to associate the long rise timescale of V1515 Cyg to a
non-triggered outbursts, and the fast rise in V1057 and FU Ori to a
triggered one. Models of triggered outbursts have been constructed,
both in the case where the outburst is artificially triggered
\citep{clarkelin90,belletal95}, and where the trigger is provided by
the interaction between the disc and an embedded massive planet
\citep{clarkesyer96,LC04}.

The environment of FU Orionis objects shows further complexities: there
is, in fact, substantial evidence for the presence of an infalling
envelope, a bipolar jet and strong mass loss in the form of a disc
wind. 

The best evidence for the presence of an envelope is represented by the
mid and far infrared SED, that shows a significant excess with respect
to the predictions of standard accretion disc models
(\citealt{kenyon91}, henceforth KH91). In many cases, even the
inclusion of reprocessing of the inner disc luminosity by an outer,
flared disc is not sufficient, because it would require too large
degree of flaring. Even though different possible explanations for this
long-wavelength excess have been discussed \citep{LB2001,abraham04}, it
is commonly attributed to reprocessing from a dusty infalling envelope,
whose inner radius should be located at $\approx 10$ au from the
central source \citepalias{kenyon91}. This model however, requires the
presence of a cavity in the envelope through which the central optical
source is seen. This cavity can be produced by a bipolar jet (typically
observed also in many T Tauri stars). The presence of spatially
separated knots in the jets of many Herbig-Haro objects has been indeed
interpreted as due to recurrent FU Orionis activity in young stellar
objects \citep{reipurth97}.

Strong winds have been observed in FU Orionis objects using a variety
of different tracers: optical and near-infrared lines
\citep{croswell87}, that show the characteristic P Cygni profile, lower
excitation lines of neutral metals and TiO bands \citep{hartmann96} and
even continuum radio observations \citep{rodriguez92}. Typical wind
velocities are observed to be in the range $300-400$ km/sec and mass
loss rates are observed to vary between different objects, being very
large for FU Ori ($\approx 10^{-5} \msunyr$) and roughly one order of
magnitude smaller for V1057 Cyg during the plateau phase and V1515 Cyg.

In addition, in some cases there is also evidence for the presence of
companions: the FU Orionis object Z CMa is a binary system
\citep{koresko91}, and recently a companion to FU Ori has been found
\citep{wang04,reipurth04}. Note, however, that this companion is
unlikely to be the cause of the outburst, as suggested by
\citet{bonnell92}, since its separation ($\approx 200$ au) is too large
to reproduce the fast rise of the outburst.

In this paper we present the result of a photometric monitoring
campaign of three well studied FU Orionis systems (FU Orionis, V1057
Cygni and V1515 Cygni) undertaken at Maidanak Observatory between 1981
and 2003. These new data, combined with historical photometric data,
are then used to provide a quantitative test to theoretical models. 

In particular, we use a thermal instability model to describe the onset
of the outburst and compare the colour evolution predicted by this
model to the observations of V1057 Cyg and V1515 Cyg. We obtain a good
match to the observations, better than what can be obtained using a
series of steady state accretion disc models, as done in the past
\citepalias{kenyon91}. As mentioned above, the long rise time and the
colour evolution of V1515 Cyg are well described by non-triggered
thermal instability models, whereas the data for V1057 Cyg require a
triggered model, where the triggering mechanism is here taken to be the
interaction with a small mass companion in the disc \citep{LC04}.

Furthermore, we present a new model of the interaction between the disc
wind and the infalling dusty envelope. The main assumption of this
model is that the wind strength scales with the luminosity of the
disc. We find that sufficiently strong winds (like that of FU Ori) are
able to blow the envelope out to large distances, while weaker winds
(like the one of V1515 Cyg) do not. For V1057 Cyg, the wind initially
is able to blow the envelope away to large distance, but as the
luminosity of the disc decreases the envelope falls back to small radii
over a timescale of $\approx$ 20 yrs. The dusty envelope provides
occultation of the central source. In this way, we are then able to
explain the diverse post-outburst photometric variability of the three
objects within a single simple model.

The paper is organised as follows. In section \ref{sec:obs} we describe
our new observations. In section \ref{sec:outburst} we show the
comparison between the observed colour evolution of V1057 Cyg and V1515
Cyg and the one predicted by disc thermal instability models. In
section \ref{sec:wind} we describe our model for the wind-envelope
interaction and in section \ref{sec:conclusion} we draw our
conclusions.

\begin{figure*}
\centerline{\epsfig{figure=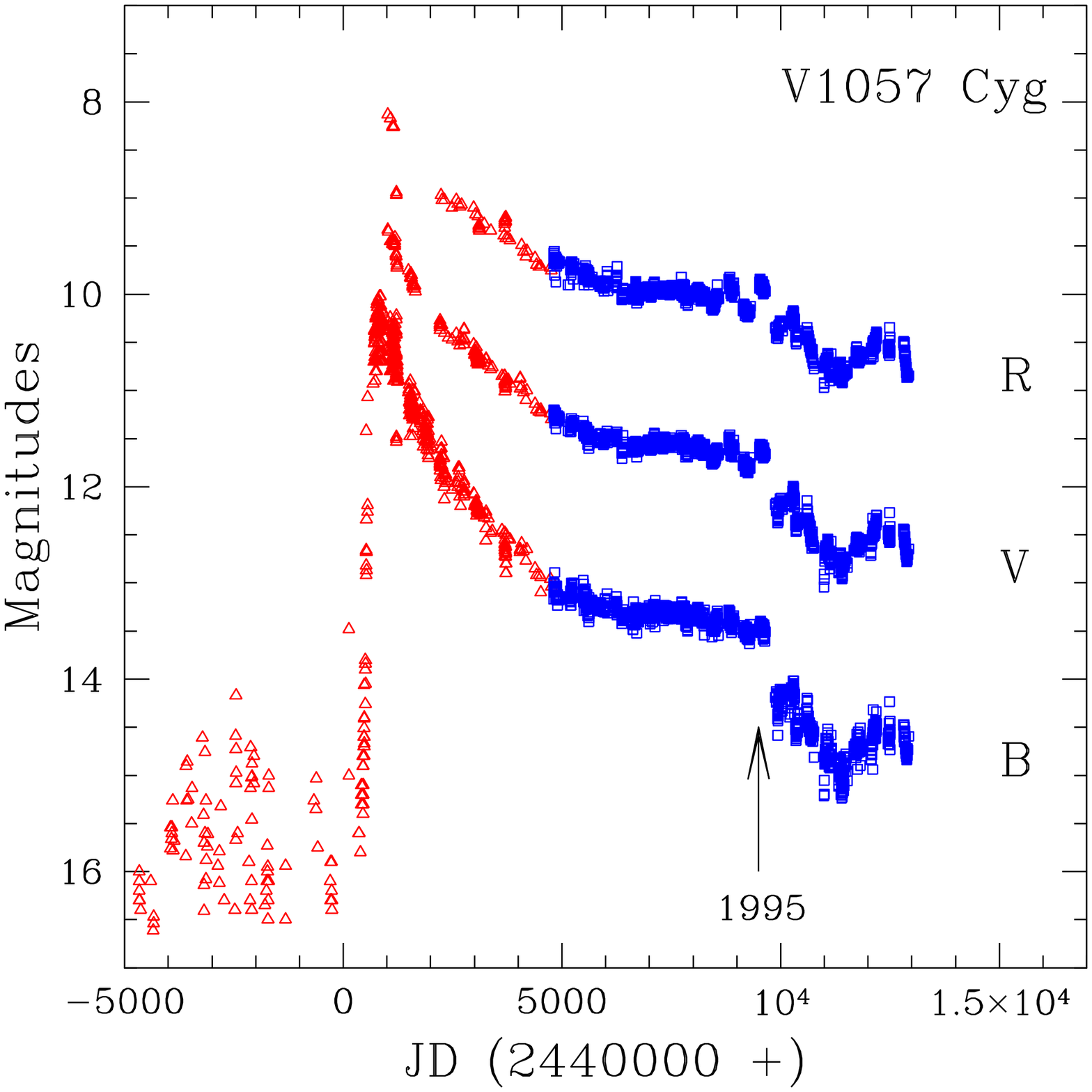,width=0.35\textwidth}
\epsfig{figure=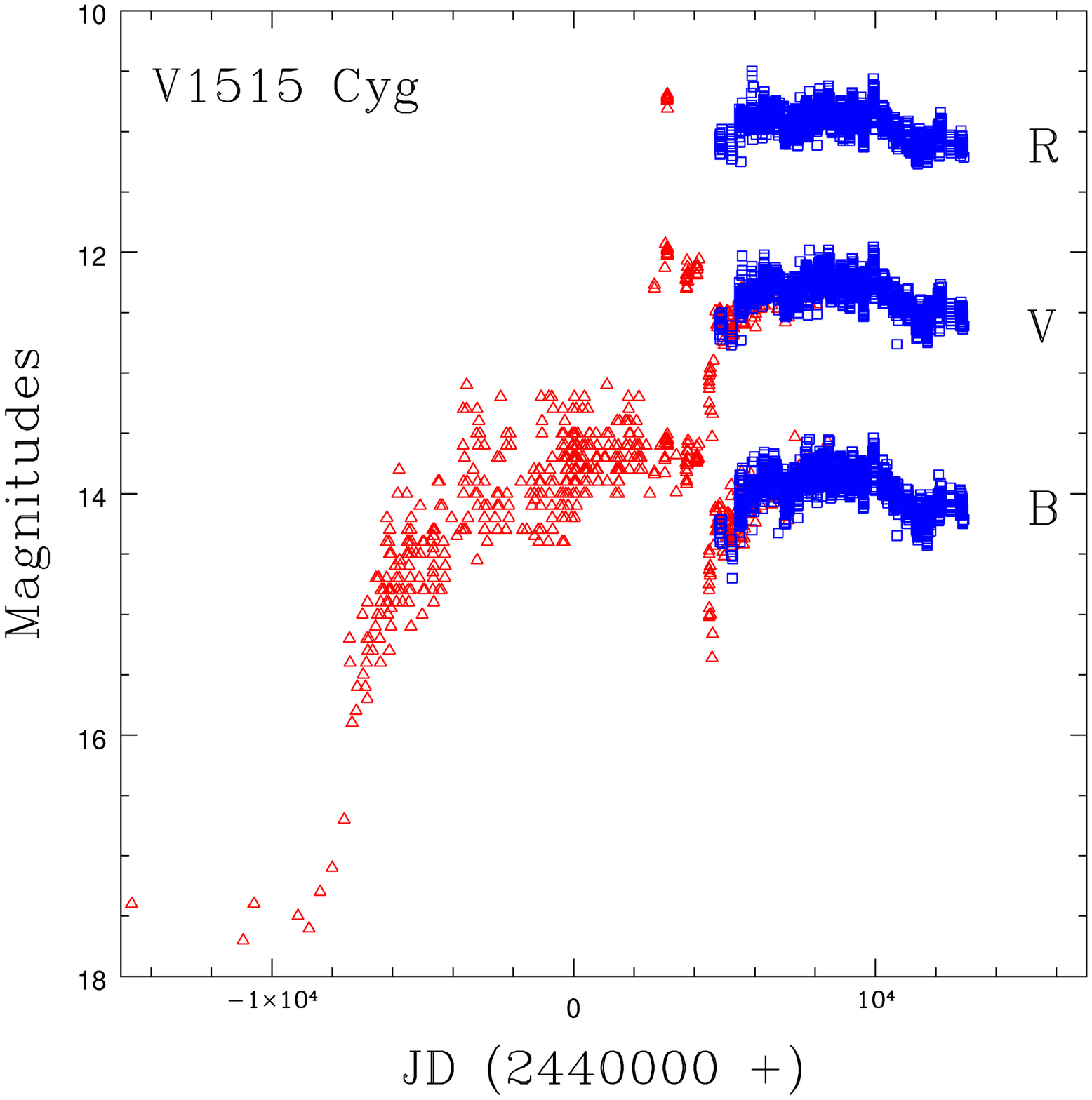,width=0.35\textwidth}
\epsfig{figure=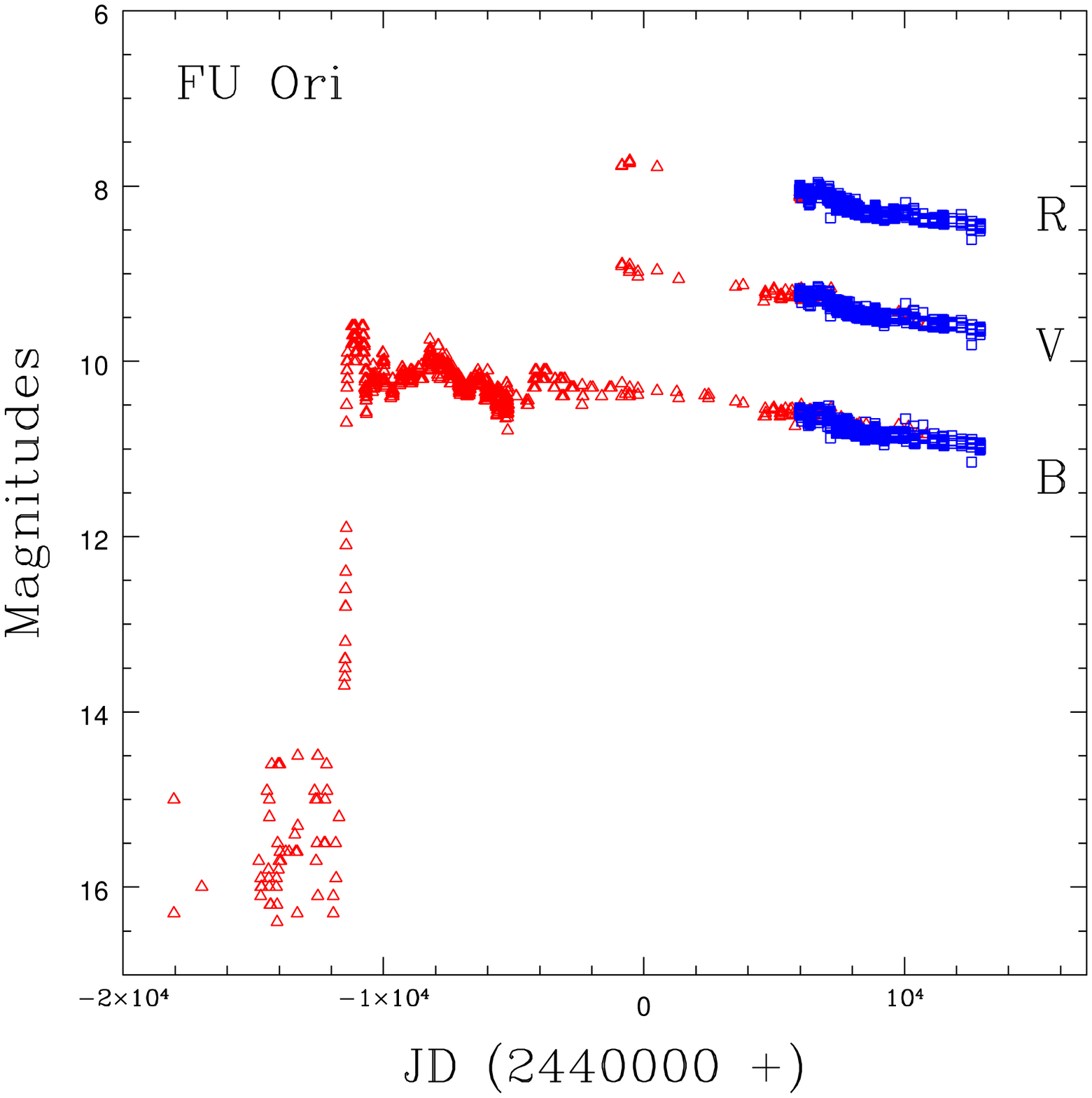,width=0.35\textwidth}
}
\caption{Light-curves of V1057 Cyg (left), V1515 Cyg (middle) and FU
Ori (right). Blue squares indicate our new data, while red triangles
indicate historical data taken from
\citet{mendoza71,rieke72,schwartz72,bossen72,welin75,welin76,
landolt75,landolt77,kolotilov77,kopa84,ibragimov88}.}
\label{fig:light_curves}
\end{figure*}

\section{Observations}

\label{sec:obs}

\subsection{New photometry}

The new $UBVR$ photometric data for FU Ori, V1057 Cyg and V1515 Cyg
presented here were taken at Maidanak Observatory during a long
monitoring campaign from 1981 (for FU Ori, from 1984) to 2003, using two
60-cm Zeiss reflectors and the 48-cm AZT-14 reflector equipped with
identical photon-counting photometers, which reproduced the
Morgan-Johnson system. The observations were carried out by
differential method using comparison stars. The rms error of a single
measurement was $\Delta V=\Delta(V-R)=0^m.015$,
$\Delta(B-V)=0^m.02$. Only for FU Ori itself the rms error in $U-B$ was
about $0^m.03$, whereas for V1057 Cyg and for V1515 Cyg this one was
much higher because of their brightness in $U$ and we used measurements
at $U-B$ only for approximate estimates. 

Full tables with the detailed photometry for the three program objects
are available electronically. Part of these data were already published
in \citet{ibrahimov96,ibrahimov99}.
 
\subsection{Long-term light curves}
\label{sec:long}

Fig. \ref{fig:light_curves} shows the long-term photometry of the three
objects considered combining historical data (red triangles, see
caption for references) and our new data (blue squares).

Among the three objects considered, the most interesting long-term
behaviour is the one of V1057 Cyg. This star, in fact, is the one which
shows the fastest time evolution: from the peak of the outburst in the
early '70s to 2003 its luminosity has decreased by almost 5 magnitudes
in $B$ and by 4 magnitudes in $V$. Its photometric behaviour after the
outburst can be divided in three phases: a period of rather steep
decrease of brightness, a ``plateau'' phase, and a period of smooth
variability after a sharp drop in luminosity in 1995. Maidanak
photometry provides a good coverage of these two last parts of the
light curve. The long-term strong evolution of the system makes it an
ideal candidate to test evolutionary outburst models. We have performed
this test by comparing theoretical models to the colour-magnitude
$V-(B-V)$ diagrams (Fig. \ref{fig:col_ev}), discussed below in Section
\ref{sec:outburst}. 

After the luminosity drop in 1995, slow year-to-year variability was
observed in V1057 (see Fig. \ref{fig:light_curves}). Apart from this
variability, V1057 also shows smooth variations within one year, but
while during the ``plateau'' phase the maximum amplitude of this
variability was about $0^m.2$, it increased to $0^m.5$ after the
luminosity drop in 1995. We subtracted long-term variability from the
lightcurve between 1995 and 2003 and analysed the residual time series
using the Starlink CLEAN algorithm (originally developed by
\citealt{roberts87}) to look for periodicity in the short-term
variations. Although we detected a period of 150 days during 1995-1999,
this period was not present in each of the individual seasons
and is, in fact, uncomfortably close to the total duration of each
of the seasonal light curves. 

V1057 Cyg shows also short-term variability and we analysed the time
series to look for short-term periodicity (of order several days) and
to test for flickering behaviour. We discuss in more detail this
variability in Section \ref{sec:short} below.

V1515 Cyg did not show any sharp decrease in luminosity during the last
years, but we can see a gradual decrease in brightness during 1996-1999
(see Fig. \ref{fig:light_curves}). Short-term luminosity drops have
been observed in 1980 and 1987 (see also \citealt{kenyonetal91}). 

During 2000-2003 FU Ori continued a slow decreasing of its brightness
(Fig. \ref{fig:light_curves}). FU Ori also did not show any sharp
luminosity drops in recent years.

For these last two objects, given their small luminosity evolution, a
time-dependent theoretical model would be less meaningful than for
V1057 Cyg. However, as an example, we have constructed a disc thermal
instability model and compared it with observations also for V1515 Cyg
and we describe it below in Section \ref{sec:outburst}. 

\subsection{Short-term variability and flickering}
\label{sec:short}

\citet{kenyon2000} have obtained some evidence for flickering -- a non
periodic brightness variability with an amplitude of roughly $0^m.2$,
on timescales of the order of a few days -- in FU Ori. We have analysed
our data for V1057 Cyg and for V1515 Cyg to look for a similar
flickering behaviour.

\subsubsection{V1057 Cyg}

\begin{figure}
\centerline{\epsfig{figure=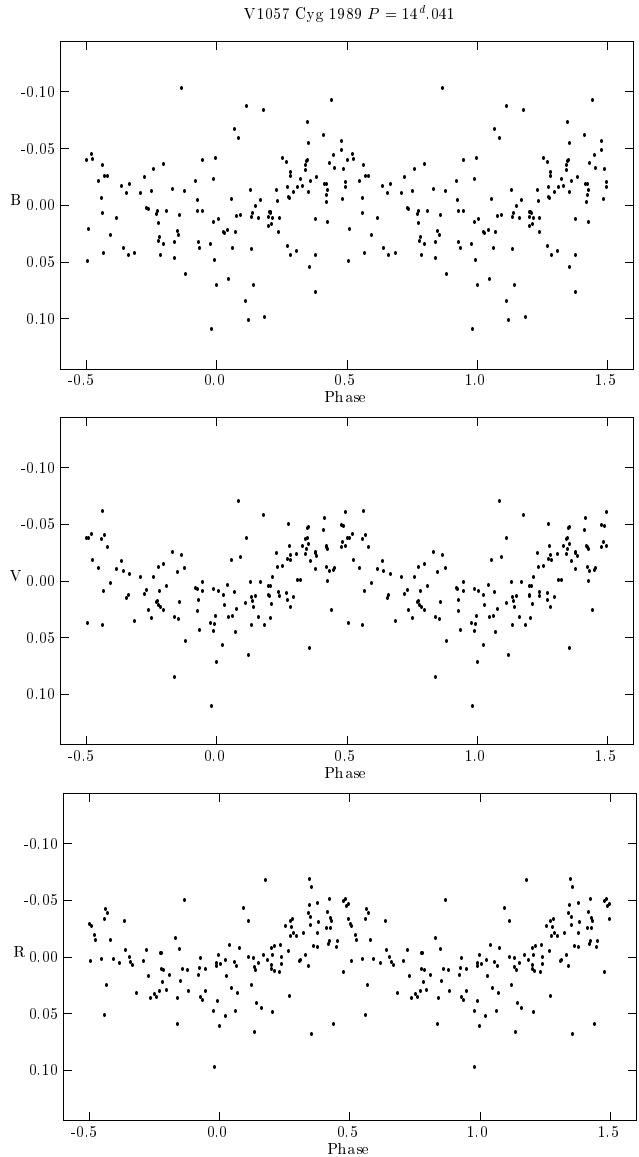,width=84mm}}
\caption{$BVR$ phased light-curves for V1057 Cyg in 1989. The period suggested
by these plots in $P\approx 14$ days.}
\label{fig:period3}
\end{figure}

We have subtracted any long-term variation from each seasonal data set
from 1986 to 1990 (i.e. during the photometric ``plateau'') by fitting
a polynomial to the observed light-curve. We have then analysed all
datasets with CLEAN. Only in 1989 (Fig. \ref{fig:period3}) our analysis
indicated the presence of a periodic component with high probability in
residual light-curve for $BVR$ data (false alarm probability smaller
than 1\%). Periodograms suggest a period of $14 \pm 0.2$ days in all
three bands, where the error represents the half width half maximum of
the periodogram peak.

\begin{figure}
\centerline{\epsfig{figure=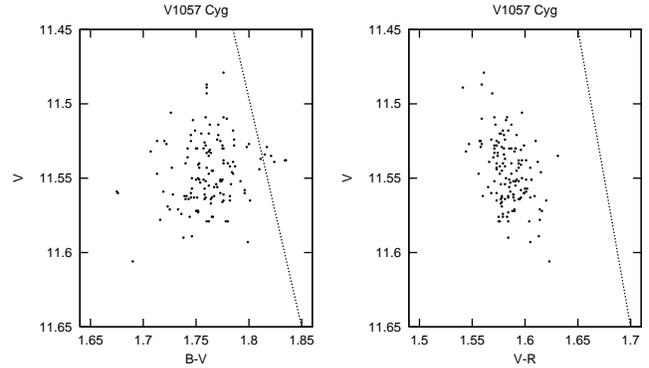,width=84mm}}
\caption{Correlation between $V$ magnitude and $B-V$ (left) and $V-R$
colours for V1057 Cyg in 1988, in which year no periodicity of the
variability was revealed by our analysis. Brightness and $V-R$ colour
variations appear to correlate well. $B-V$ colours appear to display a
larger scatter, but some correlation can still be noted. Dotted lines
show the direction of the interstellar reddening vector.}
\label{fig:flickering}
\end{figure}

For all other years, although we observe a variability with amplitude
$\approx 0^m.1$ in $V$, no periodicity was detected, as expected for a
flickering variability. To verify that these changes in brightness and
colours are real, we follow \citet{kenyon2000} and plot $B-V$ and $V-R$
colours against $V$ for 1988 (for which year no periodicity was
detected). The variations on short timescales clearly exceed
photometric errors and colour changes of $V-R$ clearly correlate with
brightness changes (Fig. \ref{fig:flickering}).  Although the scatter
in the $V-(B-V)$ diagram is bigger than in $V-(V-R)$, we can also see
some correlation between $V$ and $B-V$ (note a similar result also in
\citealt{kenyon2000} for FU Ori). Note also that the amplitude of
variation in V1057 Cyg is roughly a half of that detected by
\citet{kenyon2000} for FU Ori.

\subsubsection{V1515 Cyg}

\begin{figure}
\centerline{\epsfig{figure=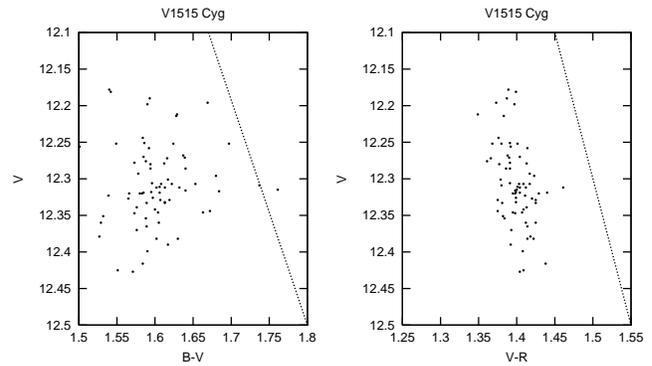,width=84mm}}
\caption{Correlation between brightness and colour variations for V1515
Cyg in 1996, during which year no periodicity in the variability was
revealed by our analysis. As for V1057 Cyg, $V-R$ colour variations
correlate well with brightness variations. Dotted lines show the
direction of the interstellar reddening vector.}
\label{fig:flickering2}
\end{figure}

We performed a similar analysis also for V1515 Cyg. After subtracting
any long-term evolution we observe a short-term variability with an
amplitude $\approx 0^m.3$ in $V$. Analysis with CLEAN only revealed
some periodicity on short time-scales from the 1987 season (although
with a slightly higher false alarm probability of 1.5 \%). As for V1057
Cyg, non-periodic colour variations in $V-R$ correlate well with
brightness variations (Fig. \ref{fig:flickering2}). Also in this case
the scatter in $B-V$ is much larger, but we can still see some
correlation in the $V-(B-V)$ diagram.

\begin{figure}
\centerline{\epsfig{figure=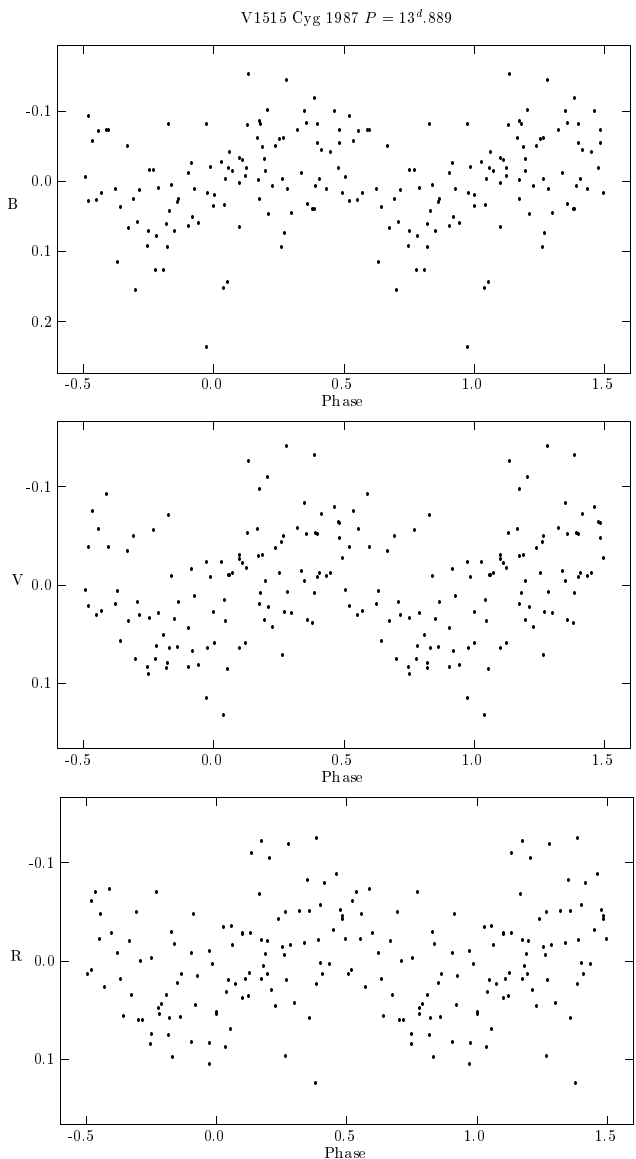,width=84mm}}
\caption{$BVR$ phased light-curves for V1515 Cyg in 1987, the only year
in which our analysis revealed some periodicity in the photometric
variability. The period suggested is $P=13.89$ days.}
\label{fig:period4}
\end{figure}

As mentioned above, our analysis indicates a periodic component with
high enough probability in residual light curves for $BVR$ data only
for the 1987 season. Periodograms suggest a period of $13.9 \pm 0.1$
days in all three bands (see phased light curves,
Fig. \ref{fig:period4}).

\subsection{Correlation between spectral and photometric variability}

FU Orionis objects show also a wealth of spectral variations on short
timescales (of the order of days). These are reviewed and discussed
extensively in \citet{herbig03}. They include spectral variations of
the equivalent width of the H$\alpha$ profiles. For example, for FU Ori
itself, during 1997-1999 \citet{herbig03} detected a variability with a
period of roughly 14 days. This kind of variability has also been
observed earlier in FU Ori by \citet{errico03}, but they detected a
smaller period of $\approx $ 6 days.

Most noticeably, \citet{herbig03} detect a periodic modulation of the
double-peaked line profiles in the cross-correlation functions of
optical photospheric absorption lines. The periods detected in 1997 are
$\approx 3$ days for FU Ori and $\approx 4$ days for V1057 Cyg (even if
the latter period is more uncertain). \citet{clarkearmi03} have shown
how a periodic modulation of these lines can be due to the presence of
a massive planet embedded in the FU Orionis discs. The observed periods
in V1057 Cyg and FU Ori would then correspond to an orbit with a
semi-major axis of roughly $10R_{\odot}$, assuming that the planet is
in Keplerian rotation around the host star. This is indeed the distance
at which we expect to find a massive planet if it is responsible for
triggering the thermal instability in the disc (see \citealt{LC04} and
Section \ref{sec:outburst} below). Note that the two objects for which
this modulation has been observed are fast-rise FU Orionis objects, for
which a triggered outburst is needed.

\begin{figure}
\centerline{\epsfig{figure=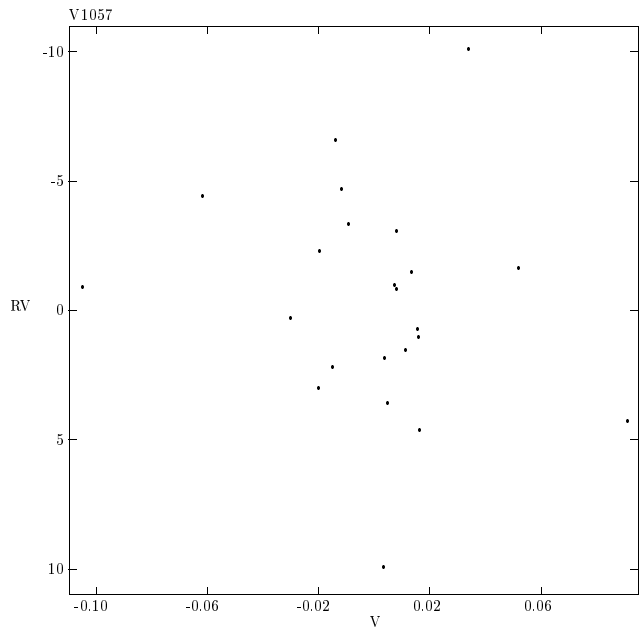,width=84mm}}
\caption{Plot of our $V$ magnitude photometry against simultaneous
radial velocity measurements by \citet{herbig03}, as derived from
variability in the shape of double-peaked photospheric absorption
lines. No apparent correlation is found.}
\label{fig:spectral}
\end{figure}

In the model of \citet{clarkearmi03}, however, the spectral variability
should not be correlated with any photometric variability. This is
because the emission pattern is in this case constant in a frame
co-rotating with the putative planet and the modulation in the line
profiles arises only from the changing location of the planet's orbital
velocity vector to the line of sight. We have then tested whether there
is any correlation between our observed photometric variability and the
modulation of the line profiles  observed by \citet{herbig03} for those periods
where our observations overlap with theirs. Fig. \ref{fig:spectral}
shows the results of this analysis for V1057 Cyg, where our $V$
luminosity is plotted against coeval radial velocity measurements by
\citet{herbig03}. No clear correlation is indeed found. A similar
result holds also for the $B$ band.

Our observations of FU Ori do not overlap with those of
\citet{herbig03}, but we used our photometry obtained from 1997 to 1999
to search for periods in the range of a few days on the light curves
(i.e. with the same periods as the observed spectroscopic period). The
analysis did not reveal any reliable photometric period in this epoch.

\section{Photometric evolution: disc thermal instability models}

\label{sec:outburst}

In this section we describe the disc thermal instability models that we
have constructed in order to describe the colour evolution of the FU
Orionis objects considered. We have considered a triggered outburst for
the fast-rise object V1057 Cyg and an untriggered one for V1515 Cyg. We
have not considered the case of FU Ori itself for two reasons: firstly
its luminosity has changed very little after the outburst (total
variation in $B$ is approximately one magnitude), secondly, as
discussed in more detail in \citet{LC04}, thermal instability models
are generally not able to reproduce such a long lived (more than 70
yrs) outburst.

\subsection{The colour evolution of V1057 Cyg}

\label{sec:inst}

V1057 Cyg is, among all the FU Orionis objects, the one that displays
the fastest decay from the outburst phase. In fact, it has faded from
the peak bolometric luminosity of $800 L_{\odot}$ to $\approx 200
L_{\odot}$ over a timescale of about 20 yrs. 

The time evolution of the Spectral Energy Distribution (SED) of V1057
Cyg has been modeled in terms of an accretion disc SED by
\citet{kenyon91}. They compared the observed SED of V1057 Cyg at
different epochs with synthetic disc models including reprocessing from
an infalling envelope, in order to account for the observed far infrared
excess. For each epoch, they assumed the disc to be in a steady state,
characterized by a constant mass accretion rate $\dot{M}$, whose value
was chosen so as to match the instantaneous luminosity of the system.
They assumed a reddening correction of $A_V=3.5$ and found an overall
good agreement with the data, except that the models tended to be
significantly redder than the observations at early epochs, i.e. close
to the luminosity maximum. However, these models are clearly limited by
the assumption that at any given epoch the disc is assumed to be in a
steady state, whereas it is in fact evolving quite rapidly. In the
following we will describe how a better agreement with observations of
the colour evolution of V1057 Cyg can be achieved when properly using
time-dependent accretion disc models. 

We have constructed time dependent 1D accretion disc models to account
for the onset and evolution of the outburst, in the framework of the
thermal instability model \citep{clarkelin89,clarkelin90,bellin94}.
Full details on the thermal instability model and on our numerical
modeling of the outburst can be found in \citet{LC04}.

V1057 Cyg is one of the fast rise FU Orionis objects. In fact, it
shows a rise timescale of the order of one year, comparable to the one
of FU Ori, but much shorter than the one of V1515 Cyg ($\approx 20$
yrs). We have shown \citep{LC04} that a fast rising light curve can be
obtained if the thermal instability, rather than being initiated at
the inner disc radius (as in \citealt{bellin94} models), is triggered
within the disc, at a radius of $\approx 10R_{\odot}$. Triggered
thermal instabilities also result in a much larger amplitude of the
outburst, with peak luminosity reaching $10^3 L_{\odot}$ (to be
compared with peak luminosities of $\approx 10^2 L_{\odot}$ obtained
in non-triggered outbursts, see \citealt{LC04}). This compares well
with the observations, since the fast rise FU Orionis objects (FU Ori
and V1057 Cyg) have indeed a larger peak luminosity with respect to
V1515 Cyg. It then appears reasonable to describe FU Ori and V1057 Cyg
in terms of triggered thermal instability models, and V1515 Cyg in
terms of non-triggered models.

Rather than assuming an {\it ad hoc} triggering mechanism, as
previously done \citep{clarkelin89,belletal95}, here we will
considered the case where the instability is triggered by the presence
of a small mass companion (a massive planet, for example) embedded in
the disc. 

The details of the planet-disc interaction assumed here, and its
inclusion in thermal instability models can be found in
\citet{LC04}. Here we briefly summarize the basic physical mechanism
responsible for the triggering. During quiescence, a massive planet (of
the order of $10M_{\mathrm{Jupiter}}$) is able to open up a deep gap in
the disc and will undergo Type II migration. However, in the case in
which the mass of the planet is large compared to the local disc mass,
\begin{equation}
M_{\mathrm{p}}\gg 4\pi\Sigma(R_{\mathrm{p}})R_{\mathrm{p}}^2
\end{equation}
(where $R_{\mathrm{p}}$ is the position of the planet and
$\Sigma(R_{\mathrm{p}})$ is the local disc density), the actual
migration timescale is longer than in normal Type II migration, and as
a consequence disc material will bank up upstream of the planet,
increasing the surface density of the disc and eventually leading to a
thermal instability. We have already shown \citep{LC04} the
effectiveness of this model in producing fast rising light curves in FU
Orionis objects.

The basic physical parameters that determine the behaviour of the
outburst are the mass of the planet $M_{\mathrm{p}}$ and the input mass
accretion rate at which the disc is fed by the envelope at large radii,
$\dot{M}_{\mathrm{env}}$. Additional parameters entering the models are
the value of the viscosity parameter $\alpha$, the mass of the central
star $M_{\star}$, and the inner radius of the disc
$R_{\mathrm{min}}$. In the following we will take $M_{\star}=
0.5M_{\odot}$, $R_{\mathrm{min}}= 5 R_{\odot}$ and (as customary in
thermal instability models) we will assume two different values for
$\alpha$ on the lower and upper branch of the thermally stable
equilibrium curves: we have taken $\alpha_{\mathrm{low}}= 10^{-4}$ and
$\alpha_{\mathrm{high}}= 10^{-3}$. We have then varied the value of
$\dot{M}_{\mathrm{env}}$ and of $M_{\mathrm{p}}$ in order to match the
observed colour evolution of V1057 Cyg. The best fit models have
$M_{\mathrm{p}}=10M_{\mathrm{Jupiter}}$ and $\dot{M}_{\mathrm{env}} =
10^{-5}\msunyr$. For this choice of parameters the outburst is
triggered at $R_{\mathrm{trig}}\approx 12R_{\odot}$, and we obtain a
fast-rise outburst with rise timescale of the order of $\approx 2$ yrs,
with a peak luminosity $L_{\mathrm{peak}}\approx 750 L_{\odot}$ and a
peak mass accretion rate of $\dot{M}_{\mathrm{peak}}\approx 6\times
10^{-4}\msunyr$. 

\begin{figure}
\centerline{\epsfig{figure=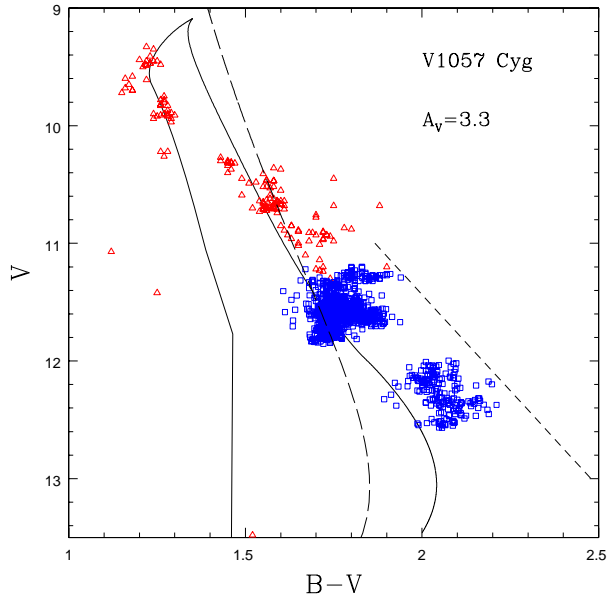,width=84mm}}
\caption{Colour evolution of V1057 Cyg. The red triangles show the
historical data points, while the blue squares show the new
observations described in this work. The solid line shows the evolution
of a thermal instability model triggered by a $10M_{\mathrm{Jupiter}}$
planet. The dashed line show the colour evolution predicted by a series
of steady state models with different $\dot{M}$. The short-dashed line
indicates the direction of the extinction vector, for a standard
interstellar extinction law.}
\label{fig:col_ev}
\end{figure}

At any given time during the outburst, the model provides us with the
full surface temperature profile of the disc, from which we can compute
the broad band fluxes by simply assuming a blackbody emission from
every annulus of the disc. Our models also include the contribution of
the central star. This contribution is, however, negligible when
compared to the high luminosity arising from the disc. The disc is
assumed to be viewed face-on and the distance to V1057 Cyg is assumed
to be $600$ pc \citep{kenyon91}. Note that (here as in the case of
V1515 Cyg, discussed below) a small inclination of $\approx
30^{\circ}$, required to fit the broadening of optical absorption
lines, can be easily accomodated by changing the assumed distance
within the uncertainties \citep{kenyon88}. The fluxes have then been
reddened assuming a reddening correction of $A_V=3.3$ (consistent with
the estimate of \citealt{belletal95}). The resulting colour evolution
in the $V-(B-V)$ plane is shown in Fig.  \ref{fig:col_ev} with a solid
line. The red triangles show the historical data, while the blue
squares show our new observations, described above. The dashed line
shows the colours of a series of steady state models, with different
mass accretion rates. These latter models are similar to those
described by \citet{kenyon91}.

Note that, for a given $V$ magnitude, close to the peak of the outburst
the time dependent models are bluer than the corresponding steady state
model. This is because in the time-dependent case, at the beginning of
the outburst, while the unstable front propagates outwards, only the
innermost part of the disc are in the high state, while the outer
parts, which contribute to longer wavelengths, are still in the low
state and therefore have a much lower luminosity with respect to steady
state model, that assumes that all the disc is in outburst. In this
way, our time dependent model is able to obtain a much better fit to
the observations at all epochs during the outburst, while the series of
steady states is redder than observed close to the peak of the
outburst, as already noted by \citet{kenyon91}.

Another thing to notice is that the simple viscous disc evolution is
not able to account for the recent drop in luminosity occurred in
1995. This is for two reasons: ({\it i}) the drop is too abrupt to be
due to viscous evolution (see light curve in
Fig. \ref{fig:light_curves}), ({\it ii}) the colour change during the
drop cannot be reproduced by the time-dependent models. In fact, all
the data points after the drop occupy a different region in the
$V-(B-V)$ plane with respect to earlier observations (see
Fig. \ref{fig:col_ev}).  The colour variation observed during the
luminosity drop is much redder than the one predicted by the viscous
evolution model. On the other hand the direction of the colour
variation is consistent with a standard interstellar extinction law,
adopting $R=3.3$ (the direction of the extinction vector is indicated
with a short-dashed line in Fig. \ref{fig:col_ev}). We are therefore
led to the conclusion that the recent drop in luminosity is due to a
sudden enhancement of the extinction along our line of sight to V1057
Cyg, rather than to the evolution of the outburst (see also
\citealt{kolotilov97}). We discuss below in section \ref{sec:wind} our
wind-envelope interaction model to explain this recent luminosity
drop. 

\begin{figure}
\centerline{\epsfig{figure=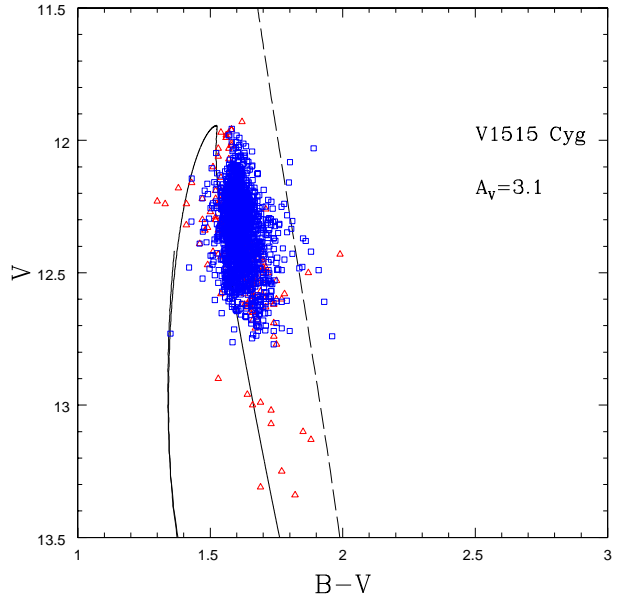,width=84mm}}
\caption{Colour evolution of V1515 Cyg. The red triangles show the
historical data points, while the red squares show the new observations
described here. The solid line shows the evolution of an untriggered
thermal instability model. The dashed line show the colour evolution
predicted by a series of steady state models with different $\dot{M}$.}
\label{fig:v1515}
\end{figure}

\subsection{The colour evolution of V1515 Cyg}

As described above, the long rise time of light curve of V1515 Cyg
suggests that in this case the thermal instability is not triggered,
but proceeds inside-out, as described by \citet{bellin94}. We have
constructed time dependent models also in this case and compared them
to the colour evolution observed in this system. In this case, we have
adopted $M_{\star}=1M_{\odot}$ and $R_{\mathrm{min}}= 3 R_{\odot}$. We
have also assumed $\dot{M}_{\mathrm{env}} = 2\times
10^{-6}\msunyr$. With these, parameters, we obtained a slow-rise
outburst, with a peak luminosity of 250 $L_{\odot}$ and a peak mass
accretion rate $\dot{M}_{\mathrm{peak}}\approx 1\times
10^{-4}\msunyr$. Fig. \ref{fig:v1515} shows the colour evolution in the
$V-(B-V)$ plane in this case. V1515 Cyg is assumed to be at a distance
of $1000$ pc, with a face-on disc. The models have been reddened
assuming $A_V=3.1$ and $R=3.3$. The assumed value of $A_V$ is slightly
smaller than the one used for V1057 Cyg, consistent with previous
estimates \citep{belletal95}. As before, the solid line shows the
result from our time dependent model, while the dashed line show the
results from a series of steady state models. The agreement is quite
good, also given the little evolution observed in V1515 Cyg.

\section{Interaction between disc wind and infalling envelope}

\label{sec:wind}

\subsection{Evidence for wind and envelope components in FU Orionis
systems} 

As mentioned above, it has long been recognised that additional
luminosity components are required in order to explain a range of
spectral diagnostics.  Here we focus on an infalling dusty envelope
(invoked by \citealt{kenyon91} in order to explain the magnitude of the
$10\mu$m excess, particularly in V1057 Cyg and V1515 Cyg) and the
strong winds in FU Orionis systems (as evidenced by P Cygni profiles
and `shell' features in optical spectra). Before proceeding to consider
the dynamical interaction between these components, we briefly describe
the observational evidence for (and the inferred properties of) each.

\citetalias{kenyon91} introduced the notion of an infalling envelope
in order to explain the magnitude of the far infrared flux (i.e. at
$\lambda > 10 \mu$m) in FU Orionis systems, which is underpredicted by
steady state, flat accretion disc models.  Although several other
suggestions have been made regarding the origin of the far infrared
emission in FU Orionis objects (see, for example,
\citealt{LB2001,abraham04}), the similar fading rates of the optical
emission and that in the range $10-25 \mu$m suggests that the latter
is derived from reprocessing of optical light (although the near
constancy of emission longward of $\sim 25 \mu\mbox{m}$ suggests an
additional component must supply the longest wavelength emission, see
discussion in \citealt{abraham04}). The dusty envelope however remains
the most likely site of emission in the $10-25 \mu$m range, since
alternative scenarios (involving reprocessing in a flared accretion
disc) are hard to construct, given the expected curvature of the disc
surface in realistic disc models \citep{belletal97}.
  
The spectral modeling of \citetalias{kenyon91} implied that the
envelope must subtend a reasonably large solid angle at the source
(covering factor $\sim 0.5$) but that the observer views the source at
rather low inclination and thus sees a low optical extinction to the
source.  This therefore implies that the envelope is somewhat
flattened (but not disc like) consistent with the expected morphology
for a rotating infalling envelope. \citetalias{kenyon91} assumed a
density profile appropriate to free fall onto a point mass

\begin{equation}
\rho = \frac{\dot{M}_{\mathrm{env}}}{4\pi r_0^{3/2}(2GM_{\star})^{1/2}}
\left(\frac{r_0}{r}\right)^{3/2}
\label{eq:dens_env}
\end{equation}
and adjusted the normalisation (and hence infall rate in the envelope)
in order that the flux reprocessed in the optically thin dust envelope
matched the spectral energy distribution longward of $10 \mu$m.  An
attractive feature of this modeling was that the inferred infall rate
($4~ 10^{-6} M_\odot$ yr$^{-1}$) is of the correct magnitude to
trigger thermal instability in the inner disc (see Section
\ref{sec:inst}). However, since the disc models can themselves provide
the observed flux shortward of $10 \mu$m (and are consistent with the
temporal behaviour of these wavebands; \citealt{abraham04}), the
envelope must not make a significant contribution shortward of $\sim
10 \mu$m.  This requirement constrains the inner edge of the envelope
to lie at around $10$ au from the central source. Since this radius is
considerably outside the dust sublimation radius (which, for a central
source luminosity of a few hundreds $L_{\odot}$, lies at a distance of
$\approx 1$ au) this begs the question of {\it what prevents the
envelope extending to smaller radii}.

  The presence of the wind is unambiguously indicated by P Cygni
profiles in Ca II $\lambda 8542$, Na I and the Balmer lines of hydrogen
\citep{croswell87,hartmann95}, indicating outflow velocities of up to
$\sim 400$ km s$^{-1}$.  Likewise, lower excitation lines of neutral
metals and TiO bands exhibit `shell features', i.e. absorption features
displaced by about $50$ km s$^{-1}$ to blueward, which are interpreted
as arising from localised condensations in the wind
\citep{herbig03,hartmann04}. Detailed modeling of these features yield
wind mass loss rates which correlate positively with the inferred
accretion rate onto the star, as is observed in other classes of young
stars \citep{hartmann95}.

\subsection{Wind-envelope interaction}

\begin{figure}
\centerline{\epsfig{figure=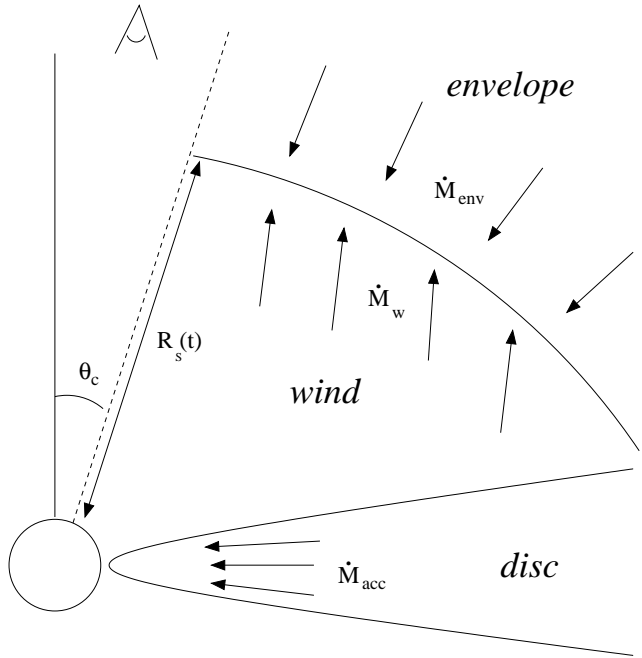,width=84mm}}
\caption{Schematic view of the system geometry assumed here.}
\label{fig:scheme}
\end{figure}

The evolution of a wind expanding into a quasi-spherical density
distribution is initially approximately adiabatic (e.g.,
\citealt{weaver77}), i.e. the combined thermal energy of the wind blown
cavity and the kinetic energy of the swept up shell of shocked medium
together increase at a rate set by the rate of kinetic energy input in
the wind ($\sim \dot M_{\mathrm{w}} v_{\mathrm{w}}^2$, where
$v_{\mathrm{w}}$ is the terminal velocity of the wind). In a flattened
density distribution, however, the wind `breaks out' in the polar
direction. We simplify the problem by considering a wind that expands
into a free-falling envelope, whose density (in spherical polar
coordinates) is given by equation (\ref{eq:dens_env}) for $\theta >
\theta_c$, where $\theta_c$ measures the angular extent of the
circumpolar cavity (see Fig. \ref{fig:scheme} for a schematic view of
the assumed geometry).  In this case, the envelope absorbs a fraction
$f= \cos \theta_c$ of the momentum of the wind. If the rate of momentum
input in the wind is $\dot p = f\dot M_{\mathrm{w}} v_{\mathrm{w}}$,
then the total momentum provided to the flow at time $t$ is given by

\begin{equation}
\label{eq:mom_int}
p(t) = \int _0^t \dot p(t') dt'.
\end{equation}

The wind drives a strong shock into the envelope, of radius
$R_{\mathrm{s}}(t)$ and the gas swept up by the shock is mainly
concentrated in a thin shell at $\sim R_{\mathrm{s}}$. For the density
distribution described in Eq. (\ref{eq:dens_env}) and assuming that the
envelope is initially in free fall (i.e with inward velocity
$v_{\mathrm{R}} = (2GM_{\star}/R)^{1/2} $ for central object of mass
$M_{\star}$), the mass contained within $R_{\mathrm{s}}$ is given by:

\begin{equation}
\label{eq:mass}
M(R_s) =
\frac{\dot{M}_{\mathrm{env}}}{3}\left(\frac{2R_{\mathrm{s}}^3}{GM_{\star}}.
\right)^{1/2} 
\end{equation}

In the limit that the gravitational acceleration of the flow is small
compared with that resulting from interaction with the wind, the
evolution of a momentum conserving thin shell is given by the
condition:

\begin{equation}
\label{eq:momentum}
p(t) = M(R_{\mathrm{s}}) \dot R_{\mathrm{s}}.
\end{equation}
  
In the simple case that $p(t)$ is a simple power law

\begin{equation}
p(t)= p_0 (t/t_0)^a, 
\end{equation}
we can readily obtain power law solutions for $R_s$, i.e.
\begin{equation}
\label{eq:similarity}
R_s = R_0 (t/t_0)^b,
\end{equation}
where
\begin{equation}
R_0=\left(\frac{15(GM_{\star})^{1/2}p_0t_0}
{2^{3/2}(1+a)\dot{M}_{\mathrm{env}}}\right)^{2/5}
\end{equation}
and $ b = 2/5(a+1)$.  

  Therefore any declining rate of momentum input implies an evolution
intermediate between the constant momentum ($R_{\mathrm{s}} \propto
t^{0.4} $) and the constant $\dot p$ ($R_{\mathrm{s}} \propto t^{0.8}$)
case.

  The neglect of the thermal and ram pressure of the material over-run
by the shock implies that this solution is valid only if $\dot
R_{\mathrm{s}}$ is greater than both the sound speed, $c_{\mathrm{s}}$
and the local free fall velocity of the unshocked envelope,
$v_{\mathrm{R}}$. Since the envelope is initially flowing in
supersonically, the latter criterion is more restrictive. The
similarity solution derived above implies $\dot R_{\mathrm{s}} \propto
R_{\mathrm{s}} ^{2a-3/2(a+1)}$, whereas $v_R \propto R_{\mathrm{s}}
^{-1/2}$. Thus $\dot R_{\mathrm{s}}/v_{\mathrm{R}}$ declines with
increasing $R_{\mathrm{s}}$ provided $a<2/3$. Therefore whereas for a
steady wind ($a=1$), gravity becomes of decreasing importance as the
shock propagates outwards, in the case of a one-off application of
momentum ($a=0$), the shell will be increasingly subject to
gravitational retardation and may be expected to re-collapse.  We shall
be applying a rate of momentum input from the wind that tracks the
luminosity evolution of the system. In the case of FU Orionis, we
adopt $a=1$, appropriate to the nearly constant system luminosity,
whereas in V1057 Cygni, the exponential decline in system luminosity
means that at late times, we expect the behaviour to approach that of a
one-off impulse.  We therefore expect the dust shell to expand
monotonically in the case of FU Orionis and for it to re-collapse in
the case of V1057 Cygni. In the latter case we may estimate the impulse
applied by the wind ($\sim \dot M_{\mathrm{w}} v_{\mathrm{w}} \Delta
t$), where $\Delta t$ is a measure of the effective timescale over
which the bulk of the momentum is applied ($\sim$ a year for this
rapidly fading system). Equating this to $p(t)$, we can therefore
obtain a rough estimate of the radius to which the shell should be
swept out (equations (\ref{eq:mass})-(\ref{eq:momentum})). We get:
\begin{equation}
R_{\mathrm{out}}=\frac{3f}{2}\frac{\dot{M}_{\mathrm{w}}}
{\dot{M}_{\mathrm{env}}} v_{\mathrm{w}}\Delta t.
\end{equation}
If we adopt the parameters we use for the simulations of V1057 Cygni
below (i.e.  $\dot{M}_{\mathrm{w}} = 6.3 \times 10^{-6} \msunyr,
\dot{M}_{\mathrm{env}} =  10^{-5} \msunyr, f=0.5$ and
$v_{\mathrm{w}}= $ 300 km/sec), we estimate that the shell should be
swept out to radii $\sim 30$ au before re-collapsing.

\subsection{Simulations}

\begin{figure*}
\centerline{\epsfig{figure=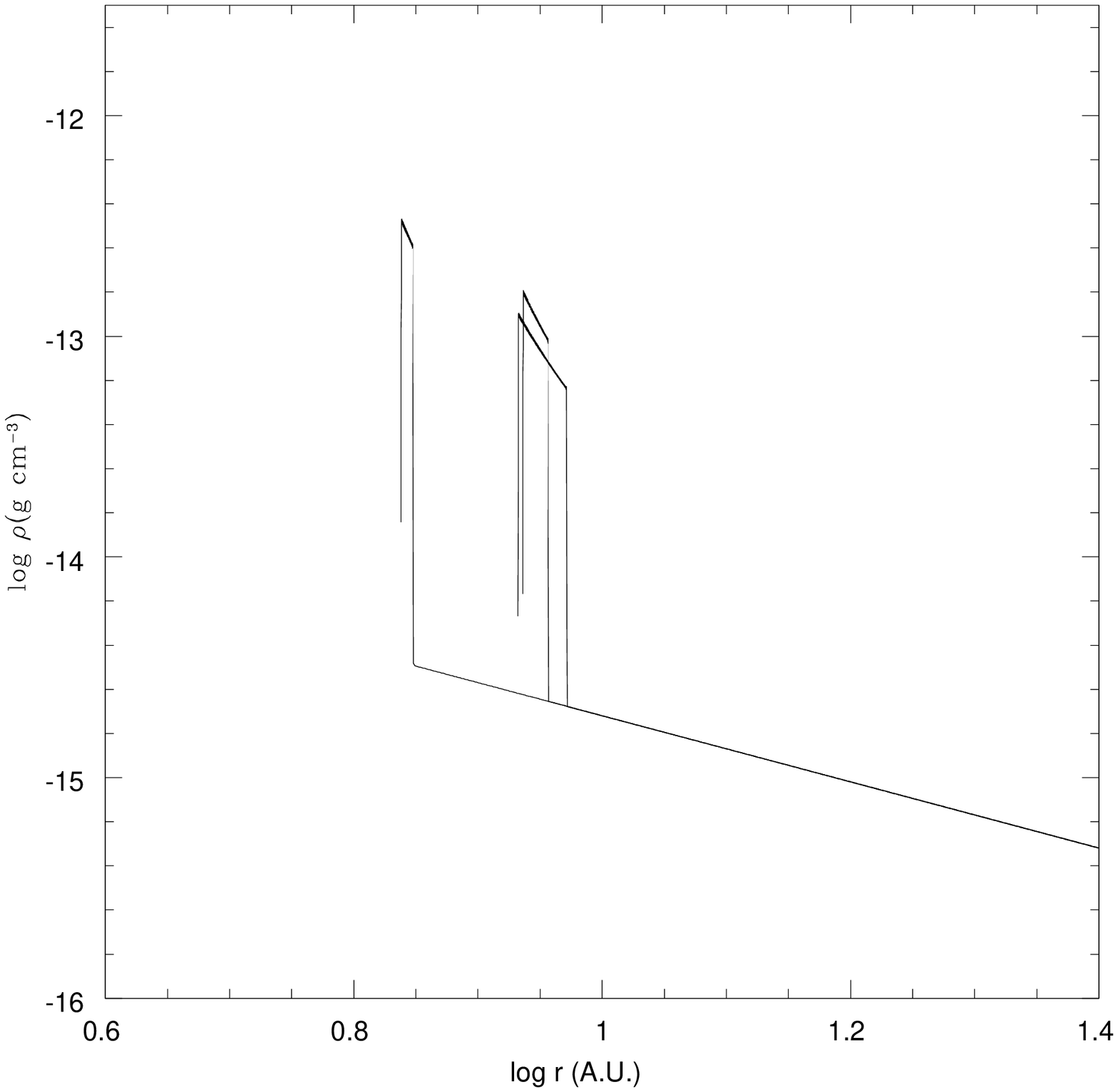,width=84mm}
            \epsfig{figure=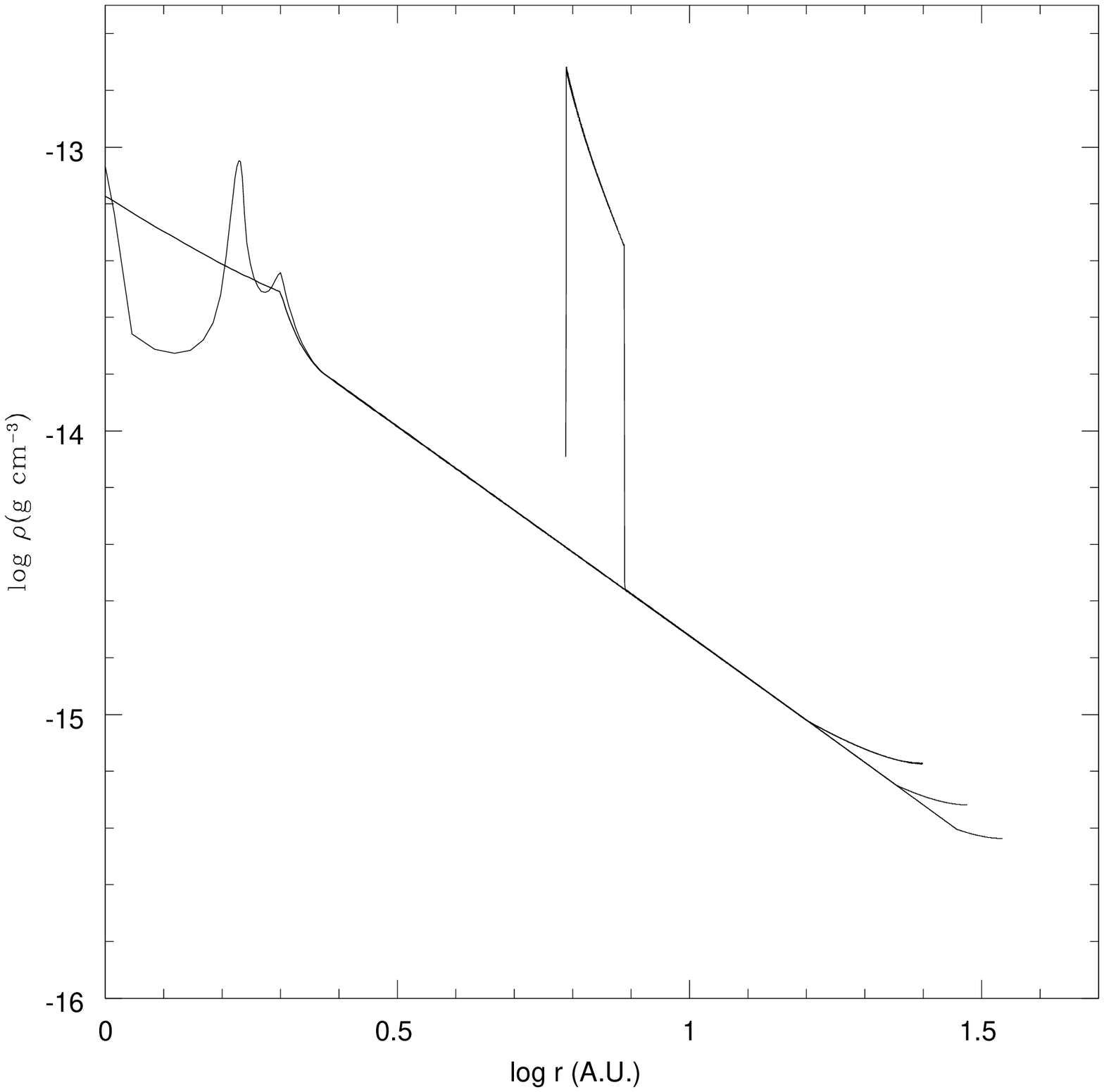,width=84mm}}
\caption{Snapshots of density evolution profile in the envelope.  The
left hand panel depicts the outward propagation of the swept up shell
(at times $4,8$ and $12$ years after the onset of outburst for
successively larger outer shell radii), while the right hand panel
shows the shell re-collapse at $16,20$ and $24$ years.}
\label{fig:wind1}
\end{figure*}

  We explore the interaction between the time dependent momentum input
from the wind and the infalling dusty envelope using a simple one
dimensional Lagrangian code which employs a standard artificial
viscosity \citep{richtmyer} in order to model the shock at the
interface between the wind and envelope.  For modeling the evolution of
V1057 Cygni, we adopt the parameters listed at the end of 4.2, and
assume a stellar mass of $M_* = 0.5 M_\odot$.  In the absence of a
proper calculation of the gas temperature in such an envelope, we
simply model the flow as isothermal gas with temperature $600$K.  A
temperature of hundreds of Kelvin is a reasonable order of magnitude
etsimate based on the expected temperature of large grains at distances
of $1-10$ au from a source with luminosity of hundreds of $L_\odot$ and
is similar to the temperature employed by \citet{muzerolle05} in
modeling the SED of the recently erupted FU Orionis system V1647
Ori. We note that the temperature does not appear in the equations
above for the evolution of a thin shell, since the only requirement
here is that the shock is strong, a condition that is readily met (with
this choice of temperature) provided the shock remains within $\approx
100$ au. We will show below that doubling the temperature changes the
outermost propagation radius and associated timescale by less than $30
\%$, so our conclusions about the applicability of this model to FU
Orionis systems are not unduly sensitive to our crude treatment of the
thermal structure of the flow.

  The impulse is initially applied to fluid elements outward of radius
$R_{\mathrm{in}}$, smoothed over a half Gaussian with standard
deviation $\sigma = 0.1 R_{\mathrm{in}}$. Since we expect the envelope
to intercept the disc at small radius, $R_{\mathrm{in}}$ is a measure
of the radius at which the envelope starts to subtend a significant
solid angle as seen from the source, and therefore corresponds to the
region where the initial wind-envelope interaction occurs. In what
follows, we adopt, somewhat arbitrarily, $R_{\mathrm{in}} = 2$
au. We note that the actual value of $R_{\mathrm{in}}$ is not
critical in determining the late time evolution of the system, unless
it is so small that the ram pressure of the envelope exceeds the
initial momentum input rate from the wind. In practice, for typical
parameters for the winds in these systems, a quasi-spherical envelope
would have to extend to radii $< 0.1$ au in order to be able to
crush the wind in this way.

  Fluid elements inwards of $R_{\mathrm{in}}$ are rapidly accreted, so
that the density distribution develops an inner cavity that propagates
outwards behind the shock. Provided that fluid elements initially at
radii $>R_{\mathrm{in}}$ remain in this region, the wind momentum
continues to be applied to them according to the initial (Gaussian)
distribution of relative weightings. If an element initially at radius
$>R_{\mathrm{in}}$ subsequently falls within $R_{\mathrm{in}} $, it is
no longer subject to momentum input from the wind, and the set of
relative weightings are passed upstream by one fluid element.  In this
way, momentum input is always distributed in the same way over the
fluid elements that are currently at radius $>R_{\mathrm{in}}$.

  For the overall normalisation of the wind momentum input rate we
assume that this varies in proportion to the total luminosity of the
source, motivated by the general correspondence between the two
discussed by \citet{hartmann95}.  Note that although we here assume
that this also roughly applies to the evolution of individual sources,
this has not been demonstrated to be the case: although V1057 Cygni,
with its dramatic fading over its first decade in outburst, would have
been a good source with which to test this hypothesis, detailed wind
modeling was only performed in the late 1980s, when the system had
already faded by a large factor. Here we model the wind of FU Orionis
as a constant mass input rate of $6.3 \times 10^{-5} \msunyr$ and that
of V1057 Cygni as having a peak mass input rate of $6.3 \times 10^{-6}
\msunyr$, in both cases assuming a wind terminal velocity of $300$
km$/$s.  We parameterise the decline of the wind in V1057 Cygni as an
exponential decline (timescale $6.4$ years) for the first $4.5$ years
followed by a slower exponential decline (timescale $15.5$ years),
thereafter.

 In order for the results to be independent of resolution, we find
that it is necessary for no more than $20 \%$ of the total wind
momentum to be applied to a single grid cell. In practice, this
implies $\Delta R/R < 0.03$ at the inner edge; in our calculations,
the shell is driven out to radii $10-100$ au and we need to model the
flow out to $\sim 50$ au in order that the front propagation
is unaffected by the outer boundary of the flow. We ran convergence test
on our simulations, employing up to $24000$ equal mass elements.

\begin{figure*}
\centerline{\epsfig{figure=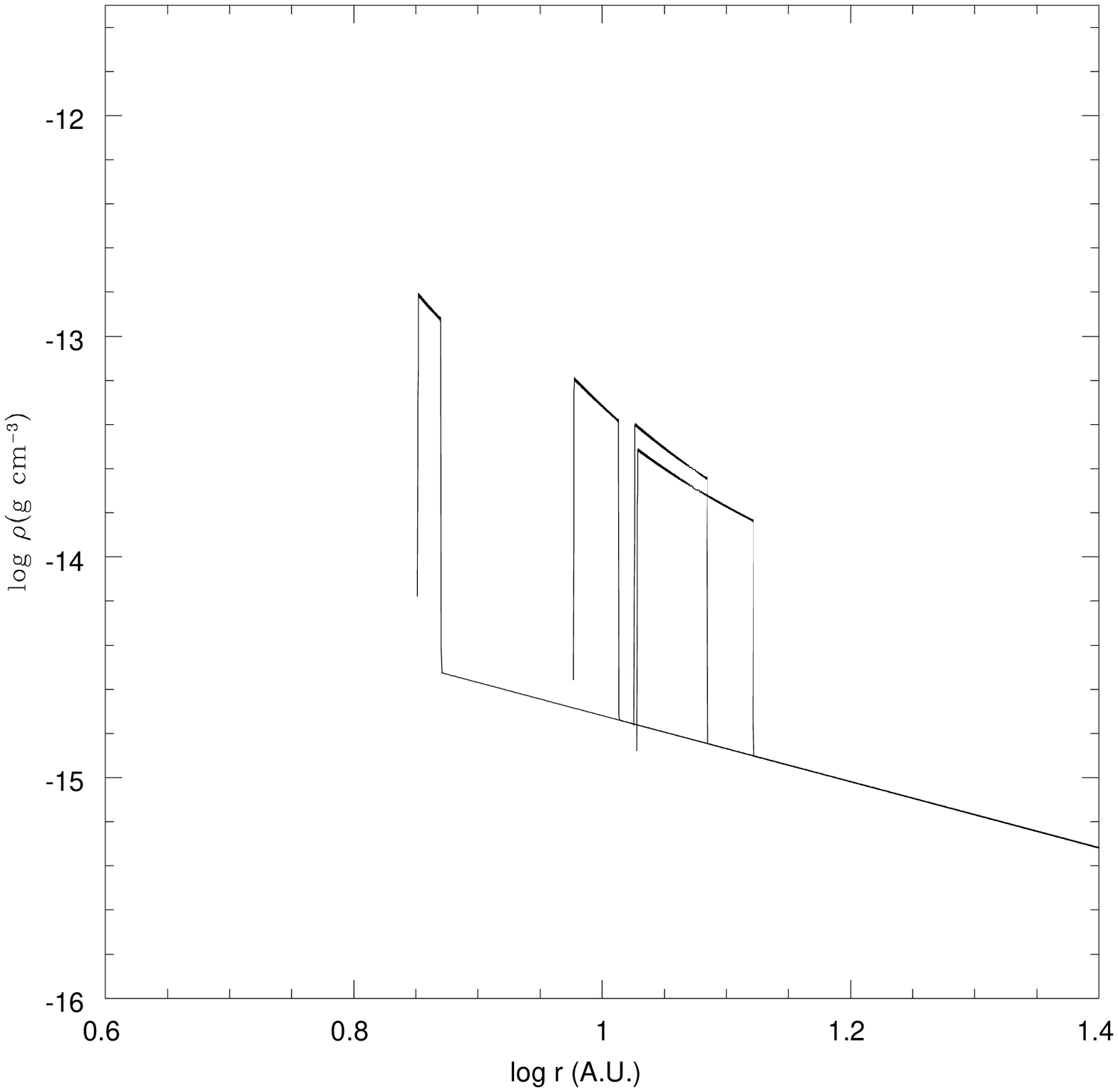,width=84mm}
            \epsfig{figure=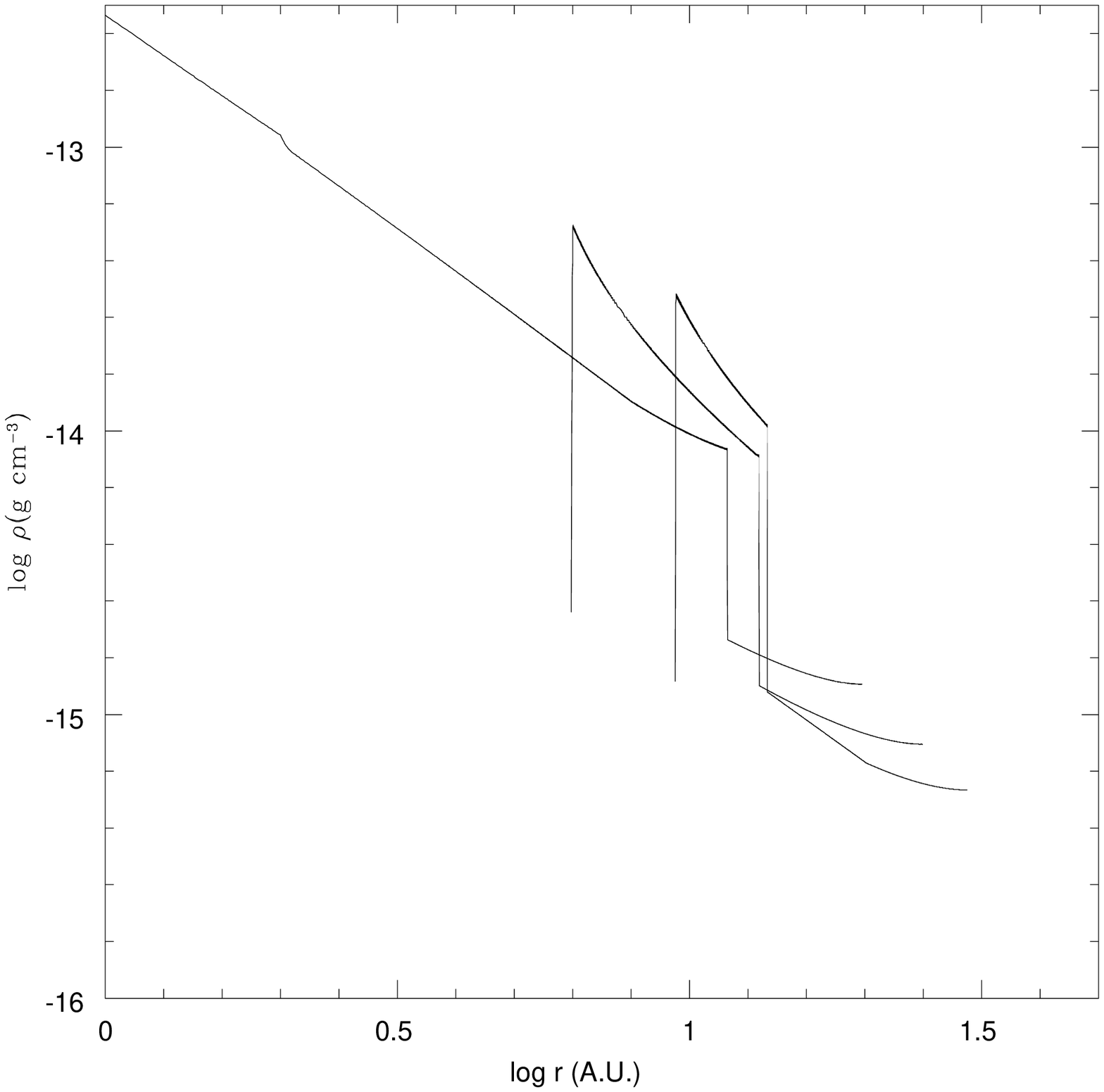,width=84mm}}
\caption{Snapshots of density evolution profile for the V1057 Cygni
simulation with $T=1200$K.  The left hand panel depicts the outward
propagation of the swept up shell (at times $4,8$, $12$ and $16$ years
after the onset of the outburst for successively larger outer shell
radii), while the right hand panel shows the shell re-collapse at
$20,24$ and $28$ years.}
\label{fig:wind2}
\end{figure*}

\subsection{Results}

  Figure \ref{fig:wind1} illustrates snapshots of the density profile
of the envelope for the V1057 Cygni models whose parameters are listed
above. As expected, the flow develops an internal cavity bounded by a
thin layer of shocked material, whose density exceeds that of the
unshocked infalling envelope by the large factor ($\sim $ the square of
the Mach number) expected for strong isothermal shocks. As expected,
given the steep decline in the momentum input from the wind, the shock
weakens as it propagates outwards and the density contrast at the shock
consequently decreases.  Gravitational deceleration causes material in
the shell to - more or less coherently - reverse direction about a
decade after outburst, when it has reached a radius of $\approx 10$ au.
The shell-like structure is initially preserved during the fallback
stage (see the density peak at $\approx 6$ au in the right hand panel
of Figure \ref{fig:wind1}, at $16$ years after outburst); the shell is
initially infalling at less than the free fall velocity and is
therefore overtaken by faster material flowing in from the envelope. As
the shell picks up speed, due to gravitational acceleration, the
velocity differential with respect to the inflowing envelope
successively decreases. Once this differential becomes subsonic, the
shock disappears and the shell-like structure is eroded, as material at
its inner edge is acclerated ahead of the rest of the flow (the remnant
shell at $\approx 2$ au is visible in the snapshot at $20$ years in the
right hand panel of Figure \ref{fig:wind1}).  By the stage that the
flow has fallen back to small radii, the density profile has relaxed to
a smooth $\rho \propto r^{-1.5}$ for free fall onto a point mass.

 We also ran a simulation identical to that above, except with a
temperature of twice the above value (i.e. $1200$K compared with
$600$K).  As one would expect, the evolution was qualitatively very
similar during the outward propagation of the shell (Figure
\ref{fig:wind2}), but differences appeared close to turn-round, where
the shell motion is temporarily subsonic. As expected, the shell is
considerably thicker in the case of the hotter simulation, and by the
time the velocity of the entire shell is reversed, the shell thickness
is already $50 \%$ of its radius (see right hand snapshot in the right
panel of Figure \ref{fig:wind2}).  The re-collapse never proceeds by
way of a thin shell (as in the colder simulation); instead, the flow
passes through a shock that remains more or less stationary in the
frame of the star on a timescale of a decade or so: the inward flowing
material is decelerated in the shock and then gravitationally
accelerated, finally free-falling onto the star. In contrast to the
colder simulation, a pronounced shock structure remains at round $10$
au, even $\approx 30$ years after the onset of outburst.
  
\begin{figure}
\centerline{\epsfig{figure=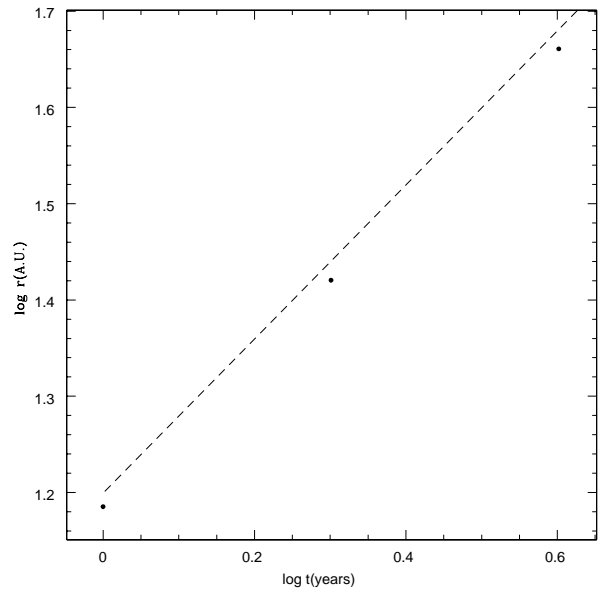,width=84mm}}
\caption{The evolution of the radius of the swept up shell for the steady wind simulation for
FU Orionis. For comparison, the dashed line depicts the $R_s \propto
t^{0.8}$ similarity solution derived in equation (\ref{eq:similarity}).}
\label{fig:windfu}
\end{figure}

  In Figure \ref{fig:windfu}, by contrast, we show how the shock radius
evolves 
for our (steady wind) model for FU Orionis.
Here, (as shown by the dashed line in Figure \ref{fig:windfu}), 
the radius of the shock evolves according to the $R_s \propto
t^{0.8}$ similarity solution derived in equation (\ref{eq:similarity})
above. As we noted above, $\dot R_s/v_R$ increases with time, so that
the role of gravity becomes less significant as the front propagates
outwards.  At the current epoch, we would expect the shock to have 
propagated to $\sim 1000$ au; if FU Orionis remained in outburst indefinitely,
then the shock would reach the outer boundary of the star's putative
natal core (at $\approx 10^4$ au) after about $10^3$ years, and would
still be strongly supersonic at this point.

\subsection{Discussion}

 We have modeled the interaction between the wind and infalling
envelope of FU Orionis systems using parameters for these components
that have previously been derived in the literature.  This modeling
has demonstrated that there are several consequences of this
interaction that can explain a number of characteristics of FU Orionis
systems. In particular, we are tempted to make the following
connections:

 i) that the inner radius of the dust envelope (invoked in order to
generate the $10-15 \mu$m flux of certain FU Orionis systems via
reprocessing of the radiation from the central source) is set by the
dynamical interaction between the wind and the inflowing envelope

  and ii) that the erratic photometric variability (observed in V1057
Cygni post 1996 and also in V1515 Cygni) is associated with the fall
back of dusty material to small radii and the consequent passage of
dust condensations across the line of sight to the inner accretion
disc.

  Below we examine each of these hypotheses in turn:

\subsubsection{Effect on the mid infrared SED}

  Figure \ref{fig:wind1} shows that in our model for V1057 Cygni, the
wind sweeps out a cavity that attains a size of $10$ au about five
years after the onset of the outburst, and which remains in the range
$10-20$ au for the next fifteen years or so.  A cavity of this
dimension is precisely what was invoked by \citetalias{kenyon91} in
their modeling of the mid infrared excess of V1057 Cygni, based on
observations made $10-20$ years after the outburst.  Our model is
therefore perfectly compatible with these results.  Thereafter, in our
model, material falls back to small radii, although, depending on the
temperature of the gas, the density discontinuity at the furthest
extent of the windblown bubble (at about $10$ au) may persist for
decades after that.  In the absence of two dimensional modeling, both
of the hydrodynamic interaction and of transfer of radiation through
the resulting structure, it is unclear to what extent we would expect
this relic structure to leave an imprint on the spectral energy
distribution.  We therefore do not attempt to predict how the SED would
evolve during this fallback phase, pending more detailed calculations.
In our models for FU Orionis, by contrast, the wind has swept the
envelope to radii $\approx 10^3$ au and we would not expect to observe
a spectral signature of a reprocessing cavity in the mid infrared.
Again, this is compatible with observations of FU Orionis, where the
evidence for extra spectral components in the mid infrared (over and
above what is supplied by a flat accretion disc) is considerably weaker
than in V1057 Cyg \citepalias{kenyon91}.

  Although we have not modeled V1515 Cyg in any detail, we note that
the weaker wind of this system (Hartmann and Calvet 1995) leads one to
expect that the envelope is not expelled to very large radii. In this
respect one would expect the system to more resemble V1057 Cyg than FU
Orionis.  In fact, V1515 Cyg is indeed similar to V1057 Cyg to the
extent that it exhibits a significant mid infrared excess.  Finally, we
note that in the recently erupted rapid rise system V1647 Ori (McNeil's
nebula), modeling of the SED derived from Spitzer and ground based data
again required an infalling envelope with infall rate of $\sim 10^{-6}
M_\odot$ yr$^{-1}$ \citep{muzerolle05}.  In this modeling, the inner
edge of this envelope was placed at $1$ au; this compact inner edge is
expected in this model a the early stage of the outburst at which this
system was modeled; depending on the subsequent luminosity evolution
during the outburst, we might expect this radius to propagate outwards
over the next few years.

\subsubsection{Effect on photometric variability}

Figure \ref{fig:wind1} shows that after around 20 years the dusty
envelope in our model for V1057 Cygni has collapsed back to small radii
and it is obviously tempting to associate this event with the sudden
dimming of the system in 1996 and its subsequent erratic variability.
(Note that, following \citealt{whitworth97}, we can readily demonstrate
that, given the densities encountered in these simulations, the dust
should not be destroyed in the shock and should instead melt only when
it falls back within the dust sublimation radius at less than an A.U.).
As has been noted (\citetalias{kenyon91}; see also Figure
\ref{fig:col_ev}, Section 3) the colour variations of these
fluctuations are consistent with screening of the source with a
variable dust screen following the standard interstellar extinction law
(Mathis 1990). It is however necessary to postulate that in the
realistic flow morphology for a recollapsing, rotating envelope, most
of the envelope joins the disc at radii $>$ a fews tenths of an au, so
that only a small fraction of its mass can intercept the line of sight
to the innermost disc which produces the bulk of the optical
emission. If this were not the case (i.e. if - as in our simplified one
dimensional model - the flow just collapsed back, re-filling a wedge of
constant opening angle), the extinction through the envelope would be
immense ($A_V > 1000$) and under these circumstances the dimming of the
source would be much more dramatic than that observed ($\sim 1$
magnitude). In fact, it is a quite realistic expectation that only a
small fraction of the envelope material should be on such low angular
momentum trajectories that it falls in to such small radii, but without
a two dimensional model we cannot quantify this further.

Such a picture in some ways resembles that proposed by
\citet{kenyonetal91} for the similar photometric variations of V1515
Cygni and also for the more recent behaviour of V1057 Cygni itself
\citep{kolotilov97}. In both cases, the variations derive from variable
extinction local to the source.  However, Kenyon et al and Kolotilov
and Kenyon postulate that this dust is formed afresh in the wind, in a
manner analogous to that observed in the winds of classical novae
\citep{gehrz88}, rather than belonging to the re-collapse of a dusty
accretion flow.  If the dust is co-moving with the wind, then the
timescale for the observed variability (of order a year, see section
\ref{sec:long}) requires that, in their model, the dust be situated at
large radii ($\sim 100$ au, see \citealt{kenyonetal91}), whereas in our
model the dust is roughly in a state of free fall and therefore must be
located at around au distance scales.

 There are several points in favour of this interpretation. Firstly,
the fallback of the envelope provides an explanation of the change in
the photometric properties of V1057 Cygni after 1995. It is also
consistent with the photometric variability patterns of other FU
Orionis systems. V1515 Cygni has demonstrated similar photometric
variability patterns to those recently observed in V1057 Cygni
throughout its outburst. The wind in V1515 Cyg is relatively weak,
comparable to that measured in V1057 Cygni after its had undergone
significant fading from the peak of the outburst \citep{hartmann95}. We
therefore postulate that whereas V1057 Cygni was able to temporarily
clear its circumstellar environment in response to the strong burst of
wind activity at the onset of its outburst, the wind in V1515 Cygni has
never succeeded in completely clearing the line of sight to the
observer of dusty material and hence the system has been subject to
erratic variations throughout the outburst. In FU Orionis, by contrast,
the stronger, more sustained, wind has succeeded in clearing a large
cavity and there is no trace in its light curve of the sort of
variability exhibited in the other two systems. If, on the other hand,
the variations result from dust formed in the wind, there is no
particularly obvious reason why the wind in FU Orionis has apparently
never undergone dust condensation events, that in V1515 Cygni has been
subject to dust condensation events throughout and the wind in V1057
Cygni has made an abrupt transition between FU Orionis like and V1515
Cygni like behaviour.

  The other advantage of our interpretation is that it does not predict
any straightforward relationship between photometric variations and the
appearance of prominent shell structures in the optical and infrared
spectra. These shell structures correspond to blue shifted absorption
features at around $50-100$ km/s \citep{herbig03,hartmann04} and are
interpreted as arising from shell-like regions of enhanced density in
the outflowing wind. There has been some discussion of the hypothesis
that the shell could be the formation site of the dust associated with
the photometric variations \citep{kenyonetal91,herbig03,hartmann04}, an
idea given additional credence by the fact that the shell features in
V1057 Cygni became very strong when the system faded dramatically at
the end of its plateau phase.  (We note in passing that the dust
formation would have to be very inefficient in this case to account for
the amplitude of such variations: the inferred column density in the
shell - $\sim 10^{23}$ cm$^{-2}$; \citealt{hartmann04} - would
correspond to $100$ magnitudes of optical extinction if the dust formed
with a standard dust-gas ratio).  More importantly, however, the shell
feature would be expected to be at its strongest at minimum light of
the system (in 1999) whereas, in fact, the shell features had
essentially disappeared at this date, to re-appear a couple of years
later as the system recovered in photometric brightness. This led
\citet{herbig03} to suggest that the photometric variations instead
result from `inhomogeneities ... passing across the line of sight', a
situation that would result from the sort of clumpy fall back that we
envisage here.  In our model, although the enhanced shell features in
1996 may or may not have anything to do with the fall back of the
envelope, the dust is not generated in, nor is necessarily cospatial
with, the shells and one would not expect to be able to trace the
effect of individual shell ejection events in the light curve of the
object.
 
 The largest uncertainties in what we have sketched above concern how
the returning flow would behave in two dimensions and what is the
degree of clumping that would be expected in reality. Two dimensional
modeling would remedy the first shortcoming: one might readily
anticipate that Rayleigh-Taylor instabilities in the interface between
the infalling envelope and wind would produce a wealth of time
dependent behaviour that could be checked against the light curves of
V1515 Cygni and V1057 Cygni. However, one would also expect that much
of the fine grained structure in the light curves of these objects
would result from small scale clumping that would not be resolved in
such simulations. Certainly, HST imaging of V1057 Cygni reveals a
wealth of features (loops and arcs) in the dust distribution on a
larger scale (hundreds of au), and it does not seem unreasonable to
suggest that the material falling in to small radii would be far from
homogeneous. In the absence of a more quantitative argument, the best
observational test of our hypothesis remains the continued
spectroscopic and photometric monitoring of V1057 Cygni and V1515 Cygni
in order that our prediction (of no correlation between individual
shell ejections and the photometric variability) can be properly
assessed.

\section{Conclusions}

\label{sec:conclusion}

Understanding the FU Orionis phenomenon is an important issue in the
general context of star formation, especially if, as is commonly
thought, most protostars undergo an FU Orionis phase during their
early evolution. In particular, given their complexity and the
different components involved in the phenomenon (protostar, accretion
disc, outflows, envelope, possible small mass companions) their
represent a valuable tool to study the interaction of the protostar
with its environment.

Here, we present new $UBVR$ photometric data of the three best studied
FU Orionis objects (FU Ori, V1057 Cyg and V1515 Cyg), taken at Maidanak
Observatory between 1981 and 2003. We have analysed our photometric
data for variability on long and short timescales. We have therefore
been able to monitor the recent evolution of these systems. Whereas
V1515 Cyg and FU Ori did not show any significant change in their
long-term evolution, V1057 Cyg has undergone an abrupt drop in
luminosity in 1995 (see also \citealt{kolotilov97}). Apart from
monitoring the secular evolution of these systems, we have also checked
for the presence of variability on shorter timescales and have detected
non periodic, flickering, behaviour on a timescale of a few days both
in V1515 Cyg and in V1057 Cyg, analogous to that observed in FU Ori
\citep{kenyon2000}. Both these objects also displayed periodic
variations (period $\sim 14$ days) during one observing season in each
case.

We have then interpreted the secular evolution of these systems within
the framework of two complementary theoretical models, both of which
involving the interaction of different components of the system. 

({\it i}) The overall outburst and subsequent luminosity decline
(especially for V1057 Cyg, that displays the fastest evolution) has
been described in terms of a standard disc instability model (the
instability being possibly triggered by the interaction between the
circumstellar disc and a massive planet or a small mass companion). For
the first time, we have compared photometric data at different epochs
with {\it time-dependent} disc models, rather than with a series of
steady state models (as done by \citetalias{kenyon91}), finding a
better agreement, with respect to the latter, of the colour evolution
of the systems, particularly in the $V-(B-V)$ colour-magnitude
diagram. In particular, we have shown that we can reproduce the large
change in system colour between 1970 and 1995 which cannot be produced
by a sequence of steady state disc models (see Figure
\ref{fig:col_ev}).

({\it ii}) On the other hand, we have interpreted the strong luminosity
drops observed early during the outburst in V1515 Cyg and more recently
also in V1057 Cyg as occultation effects due to intervening dust along
our line of sight. A similar interpretation has already been proposed
\citep{kenyonetal91,kolotilov97}, but whereas these authors assume that
the dust is formed within an expanding shell in the wind, here we
consider the occultation events as the result of the interaction
between the wind and the surrounding infalling envelope. We have
constructed a simple quasi-spherical model of the evolution of a dusty
envelope subject to the momentum input from the outflowing wind. For
sufficiently strong winds, the envelope is blown out to large radii
($\approx 10$ au, consistent with the required location of the
reprocessing dust, in order to produce the mid-infrared excess observed
in some systems, see \citetalias{kenyon91}). However, at later stages
during the outburst, as the momentum carried by the wind decreases (as
is expected if the wind strength scales with the luminosity of the
disc), the dusty envelope falls back and occults our line of sight. In
this way, our model naturally explains the different photometric
variability of the three FU Orionis objects studied here in terms of
the different strength and time dependence of the disc wind. Objects
with strongest winds (FU Ori) are able to clear away the envelope and
do not show any significant luminosity drop, those with weaker winds
(V1515 Cyg) show erratic photometric variability throughout , whereas
V1057 Cyg, which has faded substantially in the last 20 years, has
started to show variability only after the (disc) luminosity has
decreased substantially. The main limitation of our model is its 1D
nature, which makes it difficult to quantify in greater detail the
amount of occulting material that will intercept our line of sight
after the envelope has fallen back to small radii. We leave this issue
to further investigations.

\section*{Acknowledgments}

We warmly thank Jerome Bouvier for independently verifying the
photometric periods reported in this paper. We thank the Referee,
George Herbig, for a prompt and helpful review. SYM acknowledges
support from the IOA Visitors Program and CJC gratefully acknowledges
support from the Leverhulme Trust in the form of a Philip Leverhulme
Prize.

\bibliographystyle{mn2e} 
\bibliography{lodato}

\end{document}